# High thermal conductivity of high-quality monolayer boron nitride and its thermal expansion

Qiran Cai,[1] Declan Scullion,[2] Wei Gan,[1] Aleksey Falin,[1] Shunying Zhang,[1] Kenji Watanabe,[3] Takashi Taniguchi,[3] Ying Chen,[1] Elton J. G. Santos[2]* and Lu Hua Li[1]*

1. Institute for Frontier Materials, Deakin University, Geelong Waurn Ponds Campus, Victoria 3216, Australia.

2. School of Mathematics and Physics, Queen's University Belfast, Belfast BT7 1NN, United Kingdom.

3. National Institute for Materials Science, Namiki 1-1, Tsukuba, Ibaraki 305-0044, Japan.

## Abstract

Heat management becomes more and more critical, especially in miniaturized modern devices, so the exploration of highly thermally conductive materials with electrical insulation and favorable mechanical properties is of great importance. Here, we report that high-quality monolayer boron nitride (BN) has a thermal conductivity ($\kappa$) of 751 W/mK at room temperature. Though smaller than that of graphene, this value is larger than that of cubic boron nitride (cBN) and only second to those of diamond and lately discovered cubic boron arsenide (BAs). Monolayer BN has the second largest $\kappa$ per unit weight among all semiconductors and insulators, just behind diamond, if density is considered. The $\kappa$ of atomically thin BN decreases with increased thickness. Our large-scale molecular dynamic simulations using Green-Kubo formalism accurately reproduce this trend, and the density functional theory (DFT) calculations reveal the main scattering mechanism. The thermal expansion coefficients (TECs) of monolayer to trilayer BN at 300-400 K are also experimentally measured, and the results are comparable to atomistic *ab initio* DFT calculations in a wider range of temperatures. Thanks to its wide bandgap, high thermal conductivity, outstanding strength, good flexibility, and excellent thermal and chemical stability, atomically thin BN is a strong candidate for heat dissipation applications, especially in the next generation of flexible electronic devices.

## Introduction

With increasing demand in miniaturization, thermal dissipation becomes critical for the performance, reliability, longevity, and safety of various products, such as electronic and optoelectronic devices, lithium ion batteries, and micro-machines. Graphene has outstanding thermal transport: at near room-temperature, the in-plane thermal conductivity ($\kappa$) of suspended graphene produced by mechanical exfoliation and chemical vapor deposition (CVD) was mostly in the range of 1800-5300 W/mK (*1-3*) and 1200-3100 W/mK (*4-10*), respectively. The electrical conductivity of graphene, nevertheless, prevents it from being directly used in many thermal dissipation applications, such as in electronics.

It is highly desirable to find electrical insulators with high thermal conductivities. It is well-known that diamond and cubic boron nitride (cBN) are the best thermal conductors falling into this category. However, these two materials are expensive to produce due to the high temperature and pressure synthesis processes. Besides, their brittleness makes them difficult to be incorporated into flexible devices. Very recently, high-quality cubic boron arsenide (BAs) with a bandgap of 1.5 eV was found to have a $\kappa$ of ~1000 W/mK (*11-13*); however, it is unlikely to be flexible either. In comparison, the in-plane thermal conductivity of highly oriented pyrolytic hexagonal BN (HOPBN) was measured to be relatively small, *i.e.* ~400 W/mK at room temperature, but the HOPBN used in this early study consisted of small crystal domains (hence many grain boundaries), defects, and dislocations (*14*).

Atomically thin BN is a relatively new form of hBN. It has a wide bandgap of ~6 eV not sensitive to thickness change (*15*) and is one of the strongest electrically insulating materials but highly flexible and stretchable (*16*). Atomically thin BN is an excellent dielectric substrate for graphene, molybdenum disulfide ($MoS_2$), and many other two-dimensional (2D) material-based electronic and optical devices (*17*). In addition, the high thermal stability and impermeability of BN sheets are useful to passivate air-sensitive 2D materials and metal surfaces (*18,19*). The use of atomically thin BN in this aspect can be further extended to the coverage of plasmonic metal nanoparticles for surface enhanced Raman spectroscopy, enabling much improved sensitivity, reproducibility, and reusability (*20*).

The thermal conductivity of monolayer (1L) BN has never been experimentally investigated, in spite of many theoretical studies (*21-25*). There have been experimental attempts on the $\kappa$ of few-layer BN; however, most of the obtained values were less than that of bulk hBN. Jo *et al.* reported the first experimentally-derived $\kappa$ of 5L and 11L BN that were ~250 and ~360 W/mK at room temperature, respectively (*26*). One year later, Zhou *et al.* used Raman spectroscopy to find that the $\kappa$ of CVD-grown 9L BN was in the range of 227-280 W/mK (*27*). Alam *et al.* and Lin *et al.* also studied the $\kappa$ of ~30-60L and few-layer CVD-grown BN (*28,29*). Wang *et al.* measured a 2L BN using pre-patterned thermometers and deduced a $\kappa$ of 484+141/−24 W/mK at room temperature (*30*). The thickness effect has only been reported by Jo *et al.* who found 5L BN had a worse heat spreading property than 11L BN (*26*). This was opposite to the trend observed in graphene that its $\kappa$ dropped from ~4000 to ~2700 and ~1300 W/mK from 1 to 2 and 4 layers, respectively (*31*).

Therefore, it is still an open question whether atomically thin BN has higher $\kappa$ values than bulk hBN, and how the thickness affects its $\kappa$. Single-crystalline and surface-clean mono- and few-layer BN samples are needed to reveal their intrinsic $\kappa$ and the thickness effect. In the case of graphene, the crystal quality and surface cleanness could dramatically affect its $\kappa$ (*6,9*). In the aforementioned studies of few-layer BN, the samples were either mechanically exfoliated from imperfect (commercial) hBN powder or grown by CVD. Furthermore, polymer transfer processes involving either poly(methyl methacrylate) (PMMA) or polydimethylsiloxane (PDMS) were used in all these studies to prepare suspended BN, which inevitably left polymer residues. These polymer residues caused strong phonon scattering in graphene as well as in atomically thin BN due to their atomic thickness (*26*). On the other hand, thermal expansion is a fundamental property of any material, which is important to material processing and application. The experimental examination of the thermal expansion coefficients (TECs) of atomically thin BN also lacks.

Here, we report for the first time the thermal conductivity coefficients, thermal expansion coefficients of high-quality single-crystalline atomically thin BN without polymer contamination. According to optothermal Raman measurements, the suspended 1L BN had a high average $\kappa$ of 751 W/mK at close-to room temperature, and therefore it was one of the best thermal conductors among semiconductors and electrical insulators. The $\kappa$ of 2-3L BN dropped to 646 and 602 W/mK, respectively. Molecular dynamic (MD) and density functional theory (DFT) simulations were used to gain insights into the thickness effect on the $\kappa$ of atomically thin BN. In addition, we experimentally revealed that 1-3L BN had negative TECs in the range of $-3.58\times10^{-6}$/K and $-0.85\times10^{-6}$ /K at 300-400 K.

## Results

We used the Raman technique to measure the $\kappa$ of high-quality and clean atomically thin BN, as in the case of graphene (*1,3-5,27*). Atomically thin BN flakes were mechanically exfoliated from hBN single crystals (*32*) using Scotch tape onto two different substrates: silicon covered by 90 nm oxide layer (SiO$_2$/Si) and 80 nm gold-coated silicon (Au/Si), both with pre-fabricated micro-wells (diameter: 3.8 μm) and connecting trenches (width: 0.2 μm). The trenches acted as vents to avoid trapped air in BN-covered micro-wells from expansion during heating. According to our previous studies, the BN sheets prepared by this method were almost free of defects and grain boundaries (*16,18*). The absence of a polymer-based transfer process prevented surface contamination that could deteriorate thermal conductivity. An optical microscope was used to locate atomically thin sheets, followed by atomic force microscopy (AFM) to determine their thickness. Fig. 1A and B show the optical and AFM images of a 1L BN with a thickness of 0.48 nm on SiO$_2$/Si. The Raman spectra of the suspended 1-3L and bulk BN are compared in Supplementary Materials, Fig. S2.

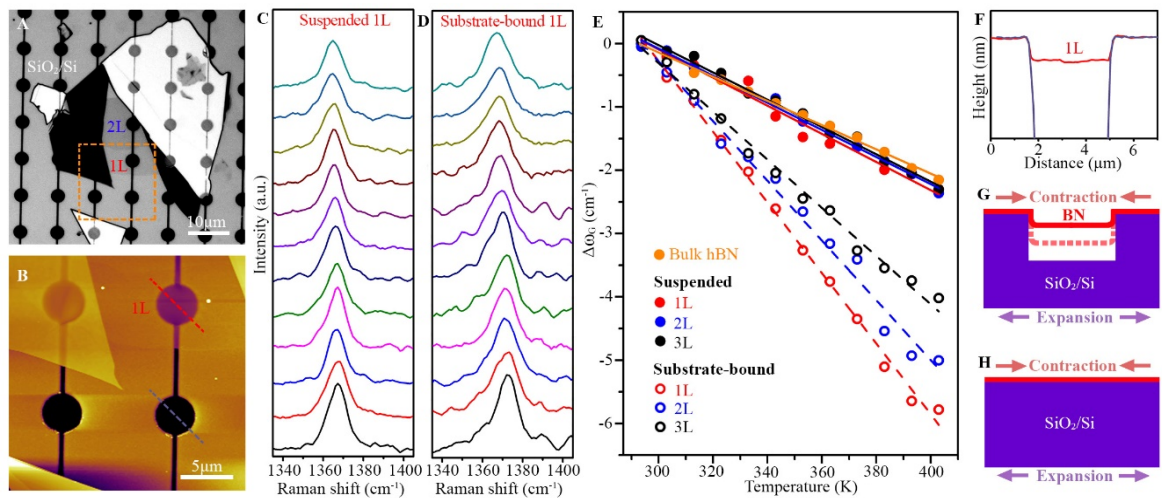

**Fig. 1. The first-order temperature coefficients.** (**A**) Optical image of a 1-2L BN on SiO$_2$/Si substrate with pre-fabricated micro-wells; (**B**) AFM image of the squared area in (A); (**C**) and (**D**) Raman *G* bands of the 1L BN suspended over and bound to SiO$_2$/Si at different heating stage temperatures from 293 to 403 K with an interval of 10 K; (**E**) summarized *G* band frequency changes of the suspended and substrate-bound 1-3L BN as a function of temperature and the corresponding linear fittings; (**F**) AFM height traces of the dash lines in (B); (**G**) and (**H**) schematic diagrams of the thermal expansion of suspended and substrate-bound BN nanosheets.

For temperature coefficients, we studied the temperature-dependent Raman spectra of three different 1-3L BN samples: suspended over $SiO_2/Si$, bound to $SiO_2/Si$, and suspended over Au/Si using relatively small laser power of 0.84-1.63 mW to minimize the heating effect. Atomically thin BN bound to Au/Si showed Raman signals too weak to be useful and hence was excluded in the study. Fig. 1C and D display the typical Raman spectra of 1L BN suspended over and bound to $SiO_2/Si$ at 293-403 K, respectively, and the Raman shifts of 1-3L and bulk hBN are summarized in Fig. 1E. Linear fittings, *i.e.* $\omega - \omega_0 = \chi T$, were applied to estimate the first-order temperature coefficients ($\chi$), where $\omega - \omega_0$ is the change of the $G$ band frequency due to temperature variation and $T$ is temperature. Interestingly, the suspended 1-3L BN showed quite similar $\chi$: $-0.0223\pm0.0012$, $-0.0214\pm0.0010$, and $-0.0215\pm0.0007$ $cm^{-1}$/K, respectively, quite close to that of the suspended bulk hBN single crystals, *i.e.* $-0.0191\pm0.0005$ $cm^{-1}$/K. In contrast, those of the substrate-bound 1-3L flakes were very different: $-0.0558\pm0.0011$, $-0.0480\pm0.0022$, and $-0.0380\pm0.0011$ $cm^{-1}$/K, respectively.

The observed frequency downshifts with increased temperature could be caused by three factors: (1) the thermal expansion of BN lattice ($\Delta\omega_G^E$); (2) anharmonic phonon-phonon effects ($\Delta\omega_G^A$); (3) the thermal expansion coefficient (TEC) mismatch between BN sheets and the $SiO_2/Si$ substrate. As atomically thin BN sheets are insulators, substrate doping was negligible (*33*). Principally, these three effects should be exactly the same for the 1-3L BN no matter suspended over or bound to $SiO_2/Si$. That is, the $SiO_2/Si$ substrate with or without micro-wells should expand the same amount with the same temperature increase. However, our results in Fig. 1E told a different story. It has been reported that mechanically exfoliated atomically thin materials, *e.g.* graphene, tend to partially adhere to the side wall of micro-wells via van der Waals attraction (*34*). Our AFM results verified the existence of this phenomenon in our suspended atomically thin BN. For example, the AFM height trace of the 1L BN in Fig. 1A and B showed that it was 29.2 nm below the surface of the substrate (Fig. 1F). We believe that the hanging-down gave the suspended atomically thin BN the capability to exempt from the third effect during heating, *i.e.* the TEC mismatch between BN sheets and the substrate. That is, the suspended atomically thin BN could peel off or adhere more to the side walls with minimum energy dissipation to fully relax and accommodate the strain owing to the TEC mismatch (Fig. 1G). This proposition was strongly backed up by our measured Raman shifts of 1-3L BN suspended over Au/Si. The TEC of Au is >30 times that of $SiO_2$ at close to room temperature, which should give rise to dramatically more influence from the TEC mismatch effect. However, the 1-3L BN suspended over Au/Si showed very similar fitting slopes to those of the samples suspended over $SiO_2/Si$ (Supplementary Materials, Fig. S3). Therefore, the intrinsic $\chi$ of 1-3L BN could be obtained from the linear fittings of the temperature-dependent Raman shifts of the BN sheets suspended over $SiO_2/Si$. The values from samples suspended over Au/Si, though close, were not used further due to the much larger TEC of Au and its potentially detrimental effect on the accuracy of $\chi$. On the other hand, the different temperature-dependent Raman shifts of 1-3L BN bound to $SiO_2/Si$ suggested different TECs of atomically thin BN, which will be discussed in detail later.

Next, the effect of laser heating on the Raman frequency of 1-3L BN suspended over Au/Si was investigated. The Au film with a much higher $\kappa$ than $SiO_2$ performed as a heat sink kept at room temperature during the measurements. Fig. 2A and B show the optical and AFM images of a 1L BN with a thickness of 0.52 nm covering four micro-wells in Au/Si. Fig. 2C-E exemplify the Raman spectra of the *G* bands of suspended 1-3L BN under different laser power. Raman downshifts were observed in all samples, suggesting increased local temperature with incremental laser power. However, such temperature increase was far from dramatic (*i.e.* ~25 K). The laser-induced Raman frequency change of the suspended atomically thin BN sheets correlated to their capabilities of thermal conduction to the edge of the micro-wells, *i.e.* heat sink. With the heat loss into the ambient (via air and radiation) in the account, the temperature distribution $T(r)$ in the suspended BN can be written as:(*4*)

$$T(r) = T_1 + \frac{Q - Q_{air}}{2\pi d\kappa} \ln\left(\frac{R}{r}\right) \beta(r), \; r \leq R \quad (1)$$

where $T_1$ is the temperature at the edge of the suspended BN, *i.e.* the boundary condition, $T(R) = T_1$ ; $Q$ is the absorbed laser power; $Q_{air}$ is the heat loss into the air; $d$ is BN thickness; $\kappa$ is thermal conductivity; $R$ is the radius of the micro-well (1.9 μm).

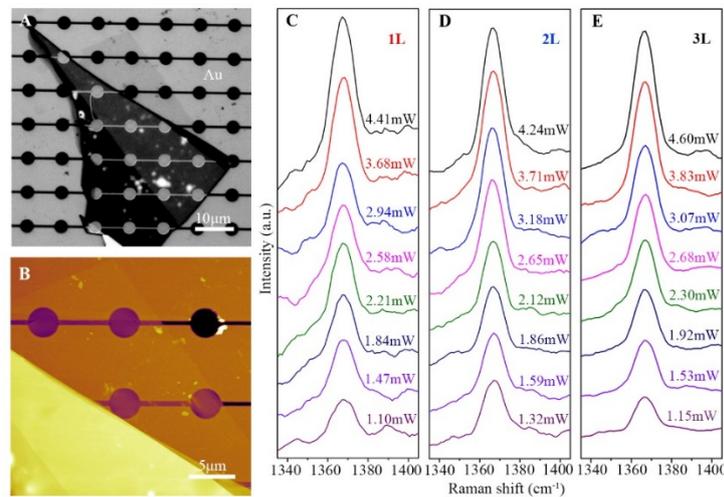

**Fig. 2. Laser power effect.** The Raman G bands of the suspended (**A**) 1L BN; (**B**) 2L BN; (**C**) 3L BN under different laser power.

It was reported that light absorption could greatly affect the accuracy of Raman-deduced thermal conductivity (*3,4*). The total laser power absorbed by the BN ($Q$) equals to the multiplication of the light absorbance ($I_{ab}$) with the laser power ($P$). We tried three methods to accurately determine the light absorbance of 1-3L BN at 514.5 nm. (1) We used PMMA to transfer atomically thin BN from SiO₂/Si onto silicon nitride (Si₃N₄) transmission electron microscopy (TEM) grids with patterned 2 μm holes. The polymer was removed by annealing at 550 °C in air. Fig. 3A-C show the optical and AFM images of a 1-2L BN before and after the transfer. The absorbance values of 1-3L BN were 0.35±0.14%, 0.62±0.19%, and 1.04±0.10%, respectively, measured by an optical power meter (Fig. 3D). These values closely followed the linear dashed line across the (0, 0) origin. (2) We also transferred atomically thin BN onto a transparent quartz plate by PMMA. The absorbance of BN sheets could be estimated by deduction of the light absorption of the quartz without consideration of the weak light reflection of 2D sheets (*35*). The absorbance values of 1-3L BN deduced from linear fitting were: 0.34±0.02%, 0.67±0.03%, and 1.01±0.04%, respectively (Supplementary Materials, Fig.

S4.). (3) We used transmitted optical microscopy under visible light, and the absorbance of 1L BN was ~0.4-0.5% (Supplementary Materials, Fig. S5), in reasonable agreement with the value measured by the optical power meter in the other two methods. For the calculation of $\kappa$, we used the absorbance values from the first method. These values were only ~15% of those measured from graphene, and much smaller than those used in previous calculations of the thermal conductivity of few-layer BN. Zhou *et al.* measured the absorbance of 1-2L and 9L CVD-grown BN transferred onto glass slides, and the values were 1.5% and 5.1% at 514.5 nm wavelength, respectively (*27*). Lin *et al.* obtained an absorbance of ~3% for a 2.1 nm-thick (6L) CVD-grown BN transferred to a quartz substrate (*29*). The small absorbance that we obtained is reasonable if one considers the wide bandgap (*i.e.* ~6 eV) of high-quality BN, which should have minimal light absorption at the wavelength of ~500 nm; however, defect states can dramatically increase its light absorption in the visible range. So the small temperature increases of the atomically thin BN with the increase of laser power (Fig. 2) were mainly due to its weak absorption of the 514.5 nm light.

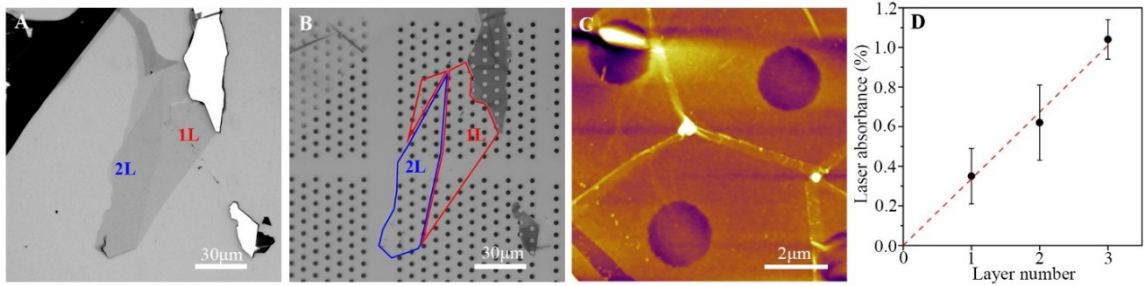

**Fig. 3. Light absorbance of atomically thin BN.** Optical images of a 1-2L BN as-exfoliated on $SiO_2$/Si (**A**) and transferred onto a $Si_3N_4$ TEM grid (**B**); (**C**) AFM image of the BN suspended over the TEM grid; (**D**) laser absorbance of 1-3L BN and the corresponding linear fitting.

$$Q_{air} = \int_{r_0}^{R} 2\pi h(T - T_a) r \, dr + \pi r_0^2 h(T_m - T_a) \quad (2)$$

where $r_0$ is the radius of the laser beam, which was estimated to be 0.31±0.01 µm by performing a Raman line scan of the edge of the Au covered Si wafer (Supplementary Materials, Fig. S6) (*4*); $h$ is the heat transfer coefficient of hBN. In the case of a small temperature difference between an object and the ambient, the quadratic expression for radiation can be linearized to reach the total heat transfer coefficient as the sum of convective($h_c$) and radiative components. Therefore, $h = h_c + \varepsilon\sigma 4T^3$, where $h_c$=3475 W/m²K for BN sheets and $\varepsilon$ is the emissivity (0.8 for hBN), and $\sigma = 5.670373\times10^{-8}$ W/m²K⁴ is the Stefan-Boltzmann constant (*36*).

$$\beta(r) = 1 + \frac{Ei\left(-\frac{r^2}{r_0^2}\right) - Ei(-\frac{R^2}{r_0^2})}{2\ln(\frac{R}{r})} \quad (3)$$

Note that $r_0$ was much smaller than the radius of the micro-wells. The Raman-measured temperature ($T_m$) of the suspended BN can be estimated by:

$$T_m \approx \frac{\int_0^R T(r) \exp(-\frac{r^2}{r_0^2}) r \, dr}{\int_0^R \exp(-\frac{r^2}{r_0^2}) r \, dr} \quad (4)$$

Given that the thermal resistance between the atomically thin BN and Au heat sink was negligible due to the relatively large contact area and high $\kappa$ of Au, the thermal conductivity of the suspended BN can be approximated as:

$$\kappa = \frac{\ln(\frac{R}{r_0})}{2\pi d \frac{T_m - T_a}{Q - Q_{air}}} \alpha \quad (5)$$

where $\alpha$ is the Gaussian profile factor of the laser beam:

$$\alpha = \frac{T_m - T_1}{T_0 - T_1} \beta(r_0) \quad (6)$$

where $T_0$ is the temperature of the suspended BN at a radial distance of $r_0$. In our experimental setup, $\frac{T_m - T_1}{T_0 - T_1}$ is ~1.03 and $\beta(r_0)$ is ~0.94, so $\alpha$ is 0.97. $T_a$ is the ambient temperature (298 K); $\frac{T_m - T_a}{Q - Q_{air}}$ denotes the increased temperature at the center of the suspended BN due to the absorbed laser power and can be deduced from Fig. 2C-E.

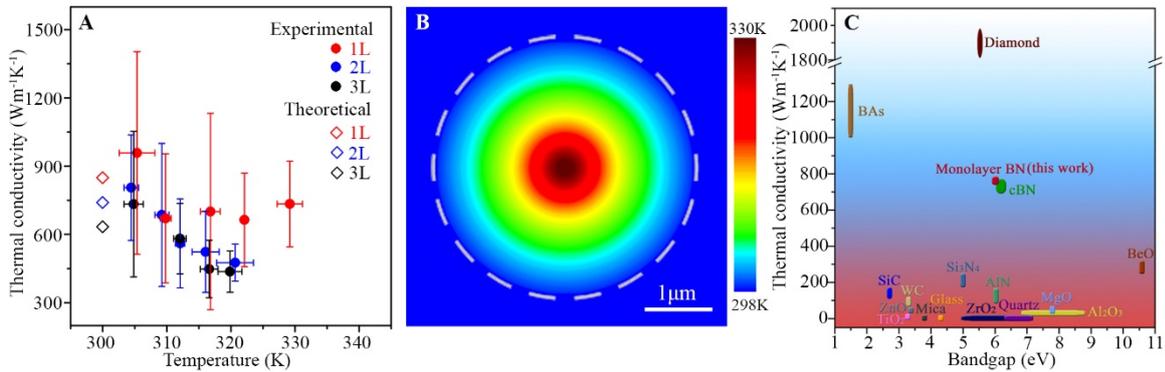

**Fig. 4. Thermal conductivity of 1-3L BN.** (**A**) Experimental $\kappa$ of the suspended 1-3L BN as a function of temperature (filled circles), and the corresponding theoretical values at 300 K (open rhombus); (**B**) the temperature distribution of a suspended 1L BN over 3.8 μm micro-wells under laser heating up to 330 K with the heat sink kept at 298 K, and the dashed line represents the edge of the suspended BN; (**C**) the comparison of the thermal conductivity of some common semiconductors and insulators.

The thermal conductivities of the suspended 1-3L BN as a function of the measured temperature $T_m$ were plotted in Fig. 4A (circles). The unusual temperature-dependent $\kappa$ of 1L BN should be due to the uncertainty in the optothermal measurements, especially its low optical absorption and hence small temperature change. Therefore, we averaged the $\kappa$ of 1-3L BN at close-to room temperature (based on totally 12 samples), and their values were 751±340, 646±242, 602±247 W/mK, respectively. It should also be noted that the optothermal method ignores nonequilibrium in different phonon polarizations, which leads to underestimated $\kappa$. The error was calculated through the root sum square error propagation approach, where the temperature calibration by Raman, the temperature resolution of the Raman measurement, and the uncertainty of the measured laser absorbance were considered (Supplementary Materials). For 1L BN, the heat loss to air ($Q_{air}$) only accounted for ~2.6% of the total heat dissipation during laser heating, and the values were even smaller for 2L and 3L BN (1.4% and 1.0%, respectively). We also used the same procedure to measure the $\kappa$ of a 1L graphene exfoliated from HOPG, which gave a value of 2102±221 W/mK (Supplementary Materials, Fig. S7). The thermal conductivities of some common semiconductors and insulators as a function of their bandgaps are compared in Fig. 4C. Monolayer BN is the third most thermally conductive semiconductors and insulators, just behind diamond and cBAs. It should also be mentioned that the thermal conductivity per unit weight of a material is important for its application *e.g.* in portable devices. hBN has a density of 2.2 g/cm³, smaller than that of cBAs (5.2 g/cm³) and cBN (3.4 g/cm³). That is, 1L BN has the second largest thermal conductivity per unit weight, just behind diamond.

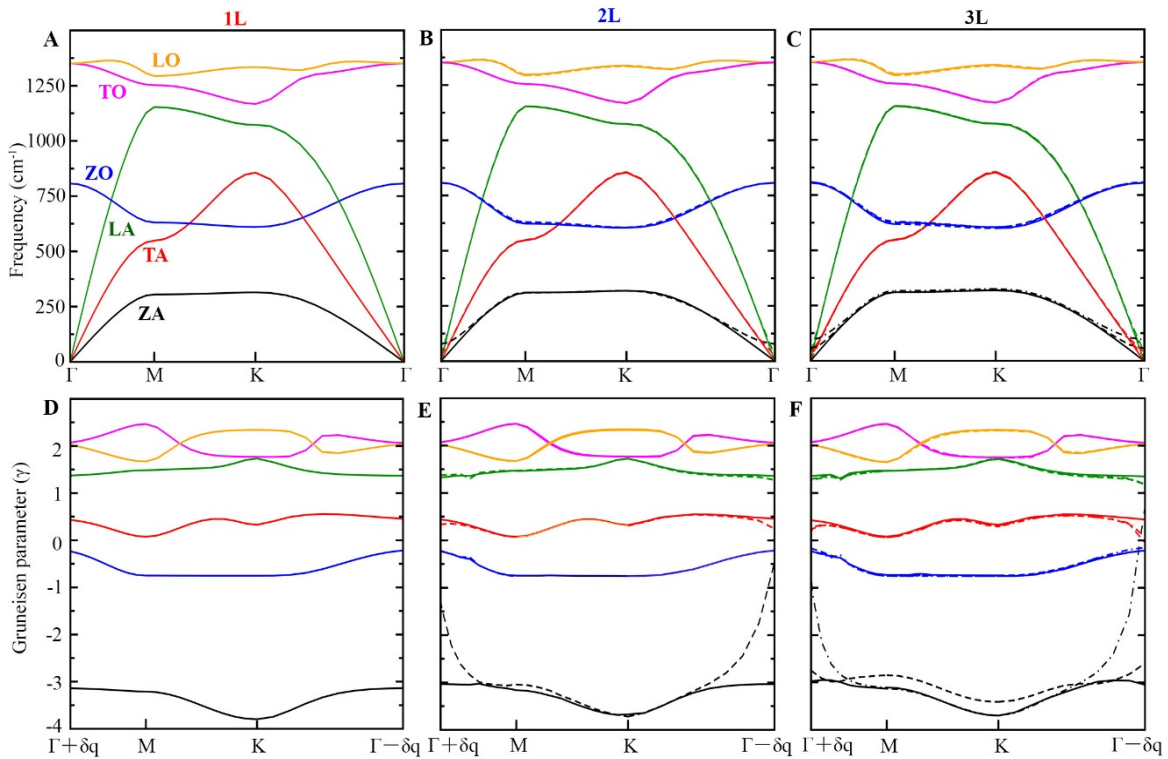

**Fig. 5. Phonon dispersion and Gruneisen parameters.** (**A-C**) Phonon dispersion and (**D-F**) Gruneisen parameters of 1-3L BN calculated by DFT. The phonon branches are labeled for 1L BN. Dashed lines represent additional phonon branches and corresponding Gruneisen parameters due to additional BN layers.

Molecular dynamics (MD) calculations (see Methods for details) were conducted on the thermal conductivity of atomically thin BN. The obtained theoretical $\kappa$ of 849, 740, and 634 W/mK for 1-3L BN at ~300K, respectively (Fig. 4A, open rhombus) were in line with the experimental values. Our calculated value of 1L BN was close to that reported by Lindsay *et al.* by considering a 10 μm grain size (*25*). The experimental trend that the $\kappa$ of BN decreased with increased thickness was also observed in the simulations. In order to explain this, the phonon dispersion and Gruneisen parameters of 1-3L BN were calculated by density functional theory (DFT) (Fig. 5). There were three optical branches, namely longitudinal optical (LO), transverse optical (TO) and out-of-plane optical (ZO) modes, and three acoustic branches, namely longitudinal acoustic (LA), transverse acoustic (TA) and out-of-plane acoustic (ZA) branches. Similar to graphene, the LO, TO and ZO branches hardly contributed to the thermal conductivity of 1-3L BN, and the ZA contribution was far larger than those from TA and LA (*31,37*). Any additional layers added to 1L BN created more ZA phonon states (Fig. 5A-C) available for Umklapp scattering, which was the dominating limitation in the thermal conductivity of defect- and grain boundary-free and surface-clean few-layer BN. Furthermore, the Gruneisen parameter and phonon frequency of the ZA mode increased with additional BN layers (Fig. 5D-F), which was further evidence that stronger Umklapp scattering occurred in few-layer BN as:

$$\frac{1}{\tau} = 2\gamma^2 \frac{k_B T \omega}{M v^2 \omega_m} \qquad (7)$$

where $\tau$ is the intrinsic phonon relaxation time associated with phonon-phonon Umklapp scattering; $\gamma$ is Gruneisen parameter; $M$ is the atomic mass; $\omega_m$ is the Debye frequency; $T$ is the temperature; $k_B$ is the Boltzman constant; and $v$ is the averaged sound velocity. Larger Gruneisen parameters and phonon frequency of ZA mode led to shorter relaxation time and more phonon scattering. The trend we observed from atomically thin BN that its $\kappa$ increased with decreasing layer thickness was also found on graphene and $MoS_2$, whose thermal conductivities increased from 1300 to 2800 W/mK, and from 98 to 138 W/mK, respectively, when their thickness decreased from 3 to 1L (*31,38-40*). It is reasonable to believe that this unusual thickness effect on thermal conductivity is intrinsic for 2D materials.

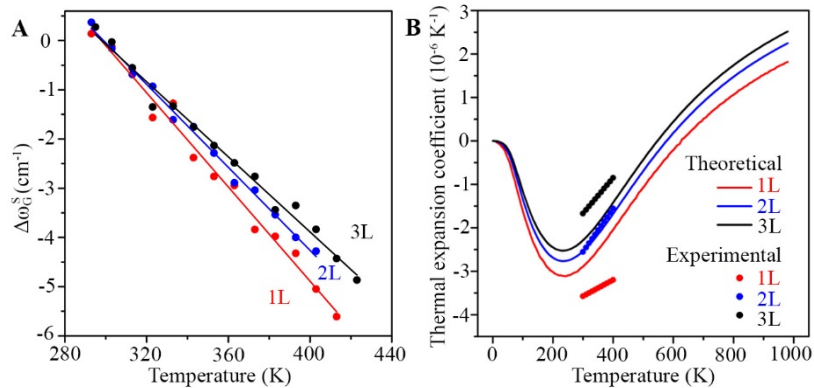

**Fig. 6. Thermal expansion coefficients of 1-3L BN.** (**A**) the *G* band frequency shifts as a function of temperature and the corresponding fittings of 1-3L BN bound to SiO₂/Si using TECs as fitting parameters; (**B**) experimental (dots) and theoretical (lines) curves of the TECs of the 1-3L BN.

As aforementioned, we could also use temperature-dependent Raman spectroscopy to estimate the TECs of atomically thin BN when it is bound to SiO₂/Si substrate, as the effect of TEC mismatch between the BN and substrate was included. The temperature-dependent *G* band shift ($\Delta\omega_G$) of substrate-bound BN nanosheets could be written as:

$$\Delta\omega_G = \Delta\omega_G^E(T_m) + \Delta\omega_G^A(T_m) + \Delta\omega_G^S(T_m) \quad (8)$$

where $T_m$ is the measured temperature of the sample; $\Delta\omega_G^E(T_m)$ is the thermal expansion of atomically thin BN; $\Delta\omega_G^A(T_m)$ is anharmonic effect; $\Delta\omega_G^S(T_m)$ is the effect of the strain $\varepsilon(T_m)$ due to the TEC mismatch between atomically thin BN and the $SiO_2/Si$, which can be expressed as:

$$\Delta\omega_G^S(T_m) = \beta\varepsilon(T_m) = \beta \int_{297}^{T_m}(\alpha_{SiO_2/Si}(T) - \alpha_{BN}(T))dT \quad (9)$$

where $\beta$ is the biaxial strain coefficient of the $G$ band of atomically thin BN. $\beta = 2\gamma\omega_o$ and $\gamma$ is the Gruneisen parameters of 1-3L BN (Fig. 5D-F), and $\omega_o$ is the strain-free $G$ band frequency (*41*). Therefore, $\beta$ values for 1-3L BN are $-56.07$, $-56.03$, and $-55.99$ cm$^{-1}$/% for 1-3L BN, respectively. $\alpha_{SiO_2/Si}$ and $\alpha_{BN}$ are the temperature-dependent TECs of 90 nm $SiO_2/Si$ and BN sheets, respectively. We used finite element method (FEM) to accurately calculate $\alpha_{SiO_2/Si}$ in the temperature range of 300-400 K (Supplementary Materials, Fig. S9). Because the temperature-dependent $G$ band shifts of the suspended BN nanosheets were contributed only by the thermal expansion of atomically thin BN lattice ($\Delta\omega_G^E$) and anharmonic effects ($\Delta\omega_G^A$), we can use Eq. 9, *i.e.* the TECs of BN nanosheets as variants, to fit the experimental data of the $G$ band shifts of the substrate-bound BN nanosheets (Fig. 6A). The TECs of 1-3L BN were estimated to be $(-3.58\pm0.18)\times10^{-6}$, $(-2.55\pm0.28)\times10^{-6}$ and $(-1.67\pm0.20)\times10^{-6}$ /K at room temperature, close to those of bulk hBN and graphene (*41,42*). The TECs of atomically thin BN were also calculated by DFT, and the theoretical and experimental curves are compared in Fig. 6B. The difference between the two could be due to the limitation of the exchange-correlation functional in representing the fundamental vibrational modes, as pointed out by us recently (*43*). Atomically thin BN has the smallest TECs among the commonly studied 2D materials (Supplementary Materials).

## Discussion

In summary, suspended high-quality and surface-clean monolayer and few-layer BN sheets were prepared to reveal their intrinsic $\kappa$. The Raman-deduced average $\kappa$ for 1-3L BN were 751±340, 646±242, and 602±247 W/mK at room temperature, respectively. The trend that the $\kappa$ decreased with increased thickness was caused by the interlayer interaction resulting in more phonon branches and states available for Umklapp scattering in few-layer BN. We also experimentally investigated the TECs of atomically thin BN: $(-3.58\pm0.18)\times10^{-6}$, $(-2.55\pm0.28)\times10^{-6}$ and $(-1.67\pm0.20)\times10^{-6}$/K for 1-3L BN at close-to room temperature, respectively. This study contributes to the knowledge system on the thermal conductivity of 2D materials and shows that atomically thin BN sheets have better thermal conductivity than bulk hBN as well as most of semiconductors and insulators, except diamond and cBAs. Along with its low density, outstanding strength, high flexibility and stretchability, good stability, and excellent impermeability, atomically thin BN is a promising material for heat dissipation in different applications.

## Materials and Methods

**Sample preparation and Raman measurement.** The trench-connected micro-wells in Si wafer were fabricated by the combination of photolithography and electron beam lithography (EBL). The depth for both the micro-wells and trenches was ~2 µm. A metal sputter (EM ACE600, Leica) was used to coat the Au film which served as a heat sink. The suspended atomically thin BN sheets were mechanically exfoliated on the Au/Si and SiO₂/Si from hBN single crystals. The optical microscope and AFM were Olympus BX51 and Asylum Research Cypher. A Renishaw inVia micro-Raman system equipped with a 514.5 nm laser was used. In all experiments, a 100× objective lens with a numerical aperture of 0.90 was used. All Raman spectra were calibrated with the Raman band of Si at 520.5 cm$^{-1}$. The laser power passing the objective lens was measured by an optical power meter (1916-C, Newport). A heating stage (LTS350, Linkam) was used for temperature control.

**Molecular dynamics using classical potentials.** Thermal conductivity coefficients $\kappa$ were calculated using the Green-Kubo approach (*44*), which was simulated by the integration of the time-dependent heat-flux autocorrelation functions via:

$$\kappa_{\alpha\beta} = \frac{1}{k_B T^2 V} \int_0^\infty \left\langle J_\alpha(t) J_\beta(0) \right\rangle dt$$

where $t$ is the time; $T$ and $V$ are the system temperature and volume, respectively; $J_{\alpha,\beta}$ are the components of the lattice heat current vector $\vec{J}$ along the $\alpha$ and $\beta$ components. $\langle J_\alpha(t) J_\beta(0) \rangle$ is the ensemble averaged heat current autocorrelation function. In this work $\alpha = \beta$ because of the symmetry of the hBN lattice along the in-plane. The heat current vector is defined as

$$\vec{J}(t) = \frac{d}{dt} \sum_i \vec{R_i} E_i = \sum_i E_i \vec{v_i} + \sum_i \frac{dE_i}{dt} \vec{R_i}$$

where $\vec{R_i}$, $v_i$, and $E_i$ are the position, velocity, and the energy of atom $I$, respectively. Calculations were carried out within MD simulations using LAMMPS. The three-body Tersoff potential (*45*) was used to treat the in-plane interactions, and a Lennard-Jones potential was used to treat the out-of-plane interactions. The parameters of these potentials have been described elsewhere (*25*). The DFT relaxed structures were used as an initial guess and then further minimized within LAMMPS. The system was then equilibrated under an NVT ensemble for 2.5 ns at 300 K. Following this, the Green-Kubo method was used to calculate the $\kappa$. Calculations were run under an NVE ensemble for 10 ns with a time step of 0.5 fs (Supplementary Materials, Fig. S8). The simulated $\kappa$ values converged within ~5.5 ns, after which the $\kappa$ magnitudes were averaged and used to determine the final $\kappa$ reported in this work. Different correlation lengths $p$ of 400, 500 and 600 ps with a sampling interval $s$ of 10 ps were used along with an effective volume of $N_x*2.50*N_y*4.33*N_L*3.33$ and 36,000 atoms in the calculation of the thermal conductivity. $N_x$ and $N_y$ are the numbers of unit cells in the x and y directions respectively, and $N_L$ is the number of layers.

**Ab initio density functional theory.** Theoretical calculations were carried out within DFT using the Vienna Ab-Initio Simulation package (VASP) (*46*). The generalized gradient approximation (*47*) along with many-body dispersion force-corrections (*48,49*) was utilized along with a well-converged 800 eV plane-wave cut-off. The projector augmented wave (PAW) (*50*) pseudopotentials were utilized to model the bonding environments of B and N. The atomic positions and lattice vectors were allowed to relax until the forces on the atoms and pressure on the cell were less than 0.000005 eV/Å and 0.01 GPa respectively. A 24×24×1 Γ-centered k-grid was used to sample the Brillouin zone. Thermal expansion coefficients were calculated using the Phonopy code (*51*), and the quasi-harmonic approximation. A 2×2×1 supercell was used in all phonon calculations.

**Acknowledgments**

**Funding：** L.H.Li thanks the financial support from Australian Research Council (ARC) via Discovery Early Career Researcher Award (DE160100796). Q.Cai acknowledges ADPRF from Deakin University. Part of the work was done at the Melbourne Centre for Nanofabrication (MCN) in the Victorian Node of the Australian National Fabrication Facility (ANFF). D.S. thanks EPSRC for studentship support. E.J.G.S. acknowledges the use of computational resources from the UK national high-performance computing service (ARCHER) for which access was obtained via the UKCP consortium (EPSRC grant ref EP/K013564/1); the UK Materials and Molecular Modelling Hub for access to THOMAS supercluster, which is partially funded by EPSRC (EP/P020194/1). The Queen's Fellow Award through the grant number M8407MPH, the Enabling Fund (A5047TSL), and the Department for the Economy (USI 097) are also acknowledged. Y.Chen acknowledges the DP150102346 and LE120100166 awards from the Australian Research Council.

**Author contributions:** The manuscript was written through contributions of all authors. All authors have given approval to the final version of the manuscript. L.H.Li conceived and directed the project. Q.Cai, W.Gan and L.H.Li prepared the samples and conducted the experiments, and A.Falin did AFM. S.Zhang did FEM simulation. E.J.G.Santos and D.Scullion did theoretical calculations. K.Watanabe and T.Taniguchi provided hBN single crystals. Y.Chen discussed the results. L.H.Li, Q.Cai and E.J.G.Santos co-wrote the manuscript with input from all authors. **Competing interesting:** The authors declare that they have no competing interesting. **Data and materials availability:** all data needed to evaluate the conclusions in the paper are present in the paper and/or the Supplementary Materials. Additional data related to this paper may be requested from the authors.

# Supplementary Materials

## 1. Optical and AFM images of atomically thin BN samples

Fig. S1 shows additional optical and atomic force microscopy (AFM) images of the mechanically exfoliated 1-3L BN suspended over pre-fabricated micro-wells.

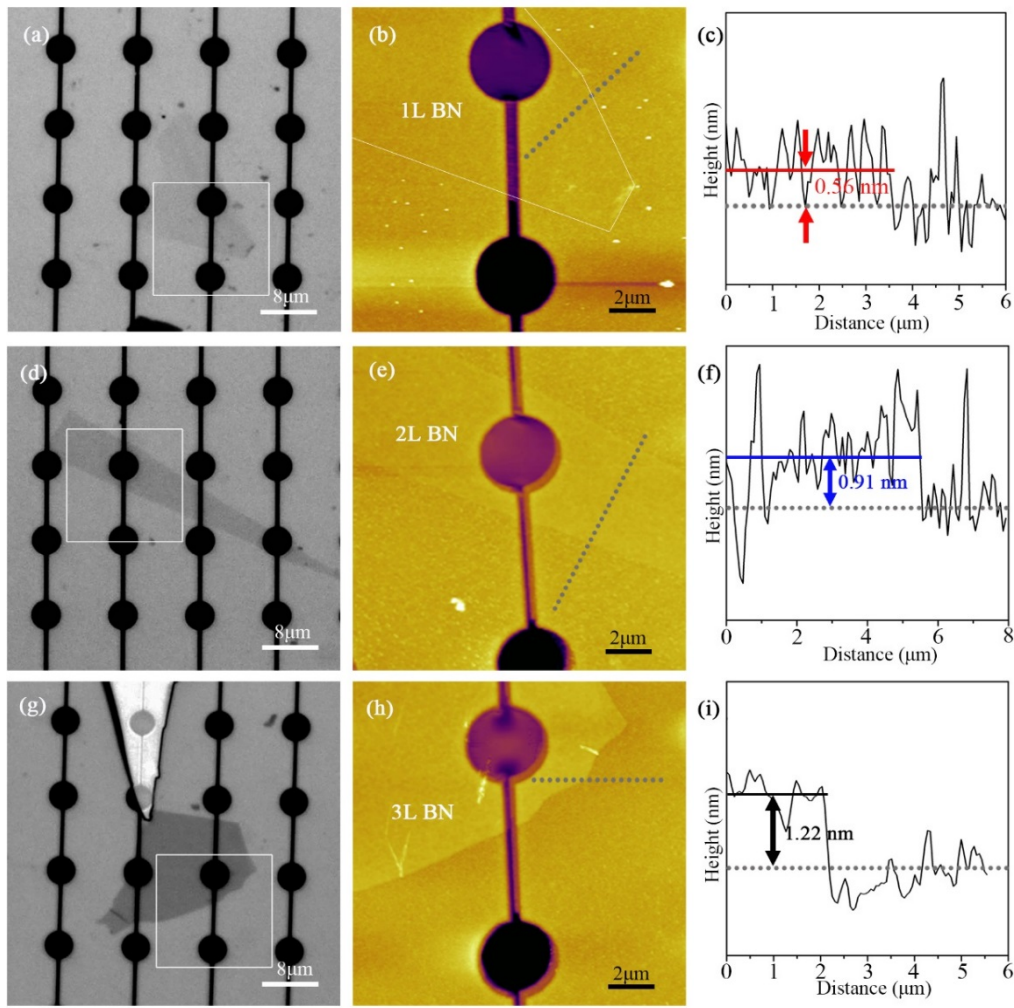

**Fig. S1.** (a-i) Optical microscopy photos, AFM images, and the corresponding height traces of additional 1-3L BN.

## 2. Raman spectra of the suspended 1-3L and bulk BN

The Raman G bands of the suspended 1-3L and bulk BN are compared in Fig. S2.

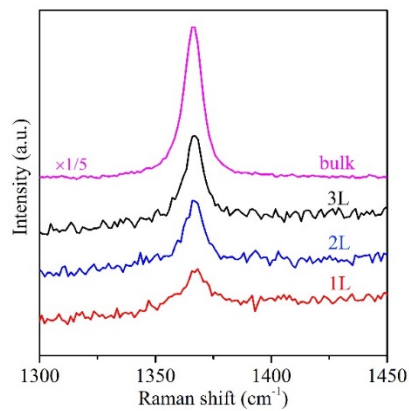

**Fig. S2.** Raman G bands of 1-3L and bulk BN.

## 3. Temperature coefficients of the 1-3L BN suspended over Au/Si substrate.

The temperature-dependent Raman G band shifts and corresponding fittings of 1-3L BN suspended over 90 nm oxide layer (SiO$_2$/Si) and 80 nm gold-coated silicon (Au/Si) are summarized and compared in Fig. S3. For Au/Si samples, the temperature coefficients of 1-3L BN were −0.0187±0.0019, −0.0196±0.0013 and −0.0199±0.0006 cm$^{-1}$/K, respectively, very close to those of 1-3L BN suspended over SiO$_2$/Si.

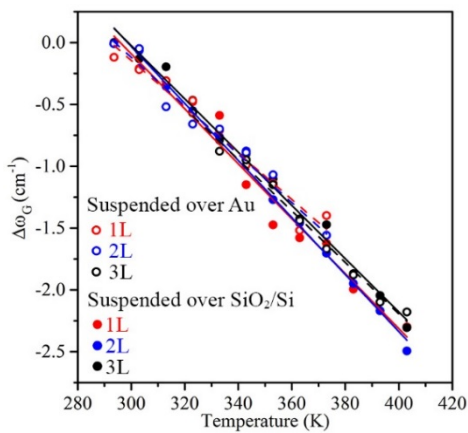

**Fig. S3.** Raman G band shifts of 1-3L BN suspended over Au/Si and SiO$_2$/Si as a function of temperature and the corresponding linear fittings.

## 4. Absorbance of 1-3L BN measured on quartz

To estimate the 514.5 nm light absorbance of 1-3L BN, we also transferred BN nanosheets mechanically exfoliated from hBN single crystals on $SiO_2/Si$ onto a transparent quartz substrate using the poly(methyl methacrylate) (PMMA) technique. To remove the PMMA, the BN nanosheets on quartz were heated at 550 °C in air for 3 h. Fig. S4a shows the reflected optical microscopy image of a 1L BN on $SiO_2/Si$. Fig. S4b and c show the digital photo of the quartz substrate and the reflected optical microscopy image of the 1L BN after transferred onto quartz and heated in air. Because of the small thickness of the 1L BN, AFM phase image was more effective than height image to visualize the transferred 1L BN on quartz (Fig. S4d and e). An optical power meter (1916-C, Newport) was then used to measure the absorbance of the BN nanosheets under 514.5 nm laser. We used UV-Vis reflectance spectrum to estimate the reflectance of the bare quartz and quartz covered by few-layer BN produced by CVD, and quartz showed a nuance (i.e. slightly higher) in reflectance with or without the coverage by few-layer BN at 514.5 nm.

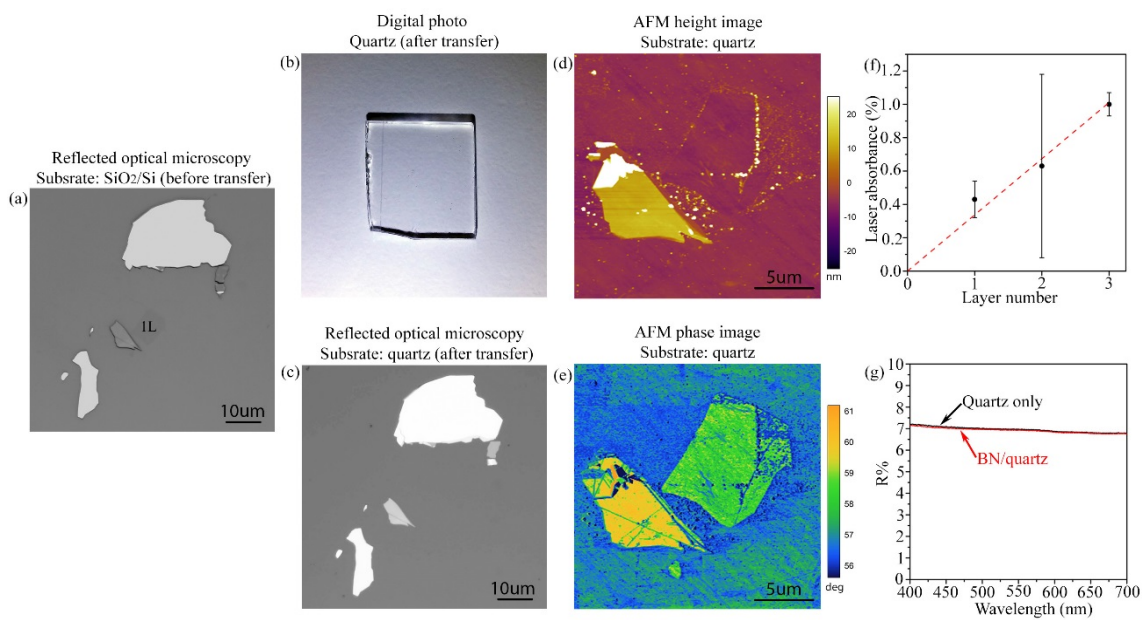

**Fig. S4.** (a) Reflected optical microscopy image of a 1L BN on $SiO_2/Si$ substrate; (b) photo of the quartz substrate after the transfer of BN and heat at 550 °C; (c) reflected optical microscopy

image of the BN nanosheets on quartz; (d, e) AFM height and phase images of the BN on quartz; (f) light absorbance of 1-3L BN and the linear fitting; (g) UV-Vis reflectance spectrum of the bare quartz and quartz covered by CVD-grown few-layer BN (BN/quartz).

We also tried to use transmitted optical microscopy to estimate the absorbance of the 1L BN under visible light. Fig. S5a shows that the 1L BN on quartz was not visible under the transmitted optical microscope, but the light intensity profiles of the dashed lines 1 and 2 did show small differences, as shown in Fig. S5a and b. The absorbance of the 1L BN estimated from the profiles was ~0.4-0.5% in the visible range, in good agreement with the value measured by the optical power meter.

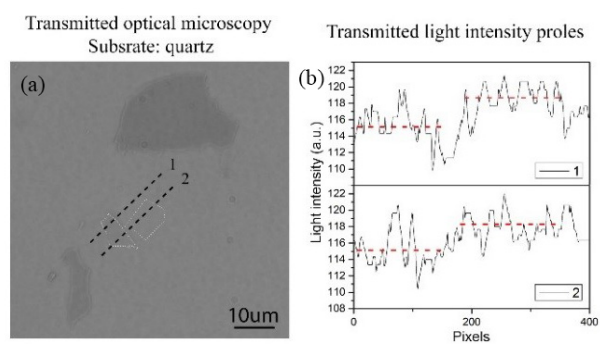

**Fig. S5.** (a) Transmitted optical microscopy image of the same BN on quartz; (b) light intensity profiles of the dashed lines 1 and 2 in (a).

## 5. Laser beam radius

The laser beam radius $r_0$ (objective lens 100×) was measured by performing a micro-Raman scan across the boundary of a partially Au-coated Si wafer. Fig. S6a and b display the optical image of the boundary with and without Au coating and the corresponding Raman mapping of the Si frequency at 520.5 cm$^{-1}$ across the boundary. The measured intensity ($I$) was proportional to the total laser power incident on the sample. The step size used in the mapping was 0.1 μm.

Fig. S6c shows the distribution of $I$ as a function of the distance ($x$) to the boundary. A Gaussian

function $exp(-x^2/r_0^2)$ was used to fit the slope $dI/dx$ to calculate $r_0$, which was found to be

0.31±0.01 μm (Fig. S6d).

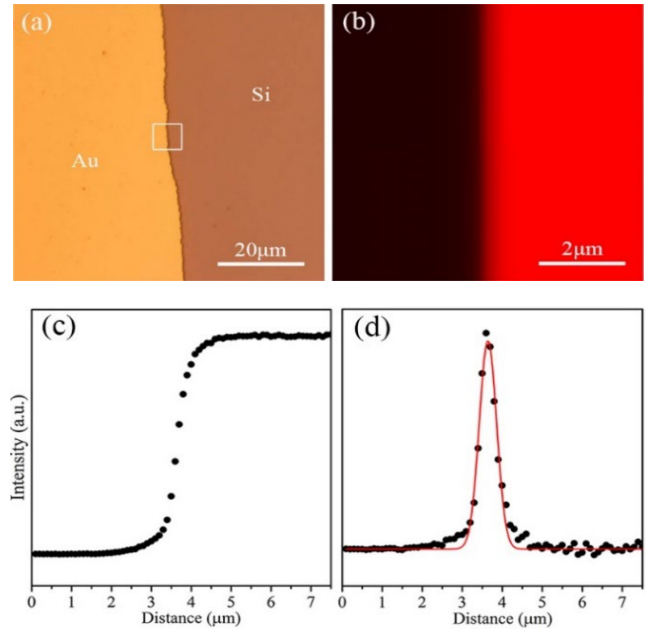

**Fig. S6.** (a) Optical microscopy image of the boundary with and without Au coating on a Si wafer used for the estimation of the laser beam radius; (b) Raman mapping of the Si band at 520.5 cm$^{-1}$ in the squared area in (a); (c) distribution of the intensity of the Si Raman band across the boundary; (d) the corresponding Gaussian fitting.

## 6. Error calculation

The thermal conductivity errors were calculated through the root sum square error propagation approach, where the temperature calibration by Raman peak position, the temperature resolution of the Raman measurement, and the uncertainty of the measured laser absorbance were considered. For example, R=1.9 μm, a=(0.35±0.14)%, $r_0$=(0.31±0.01) μm, $T_m$=(302±3)

K, P=1.1 mW, therefore $\kappa = \frac{\ln(\frac{R}{r_0})}{2\pi d \frac{T_m - T_a}{Q}} \alpha$ = ln(1.9/0.31)*0.35%*1.1*0.97*1000000/(2*3.14*

0.334*5) = 807 W/mK, and Error = $807 \times \sqrt{\left(\frac{0.01}{0.31}\right)^2 + \left(\frac{0.14}{0.35}\right)^2 + \left(\frac{3}{303}\right)^2}$ = 323, and then all the

data in the similar temperature range, for example, 807±323 W/mK at 303K, 850±300W/mK

at 300K, and 650±280 W/mK at 302 K, were averaged, and the error will be calculated by the

root sum square error propagation approach: $\kappa$ =

(807+850+650)/3±$\sqrt{(323)^2 + (300)^2 + (280)^2}$/3 = (769±175)W/mK.

# 7. Thermal conductivity of graphene as a control

For validation purpose, we used the same optothermal procedure to measure the thermal

conductivity of monolayer graphene suspended over the same substrate with 3.8 μm holes

connected by 0.2 μm wide trenches. The graphene sheets were mechanically exfoliated from

highly oriented pyrolytic graphite (HOPG) using Scotch tape. The Raman results are shown in

Fig. S7. The calculated thermal conductivity of 1L graphene was 2102±221 W/mK.

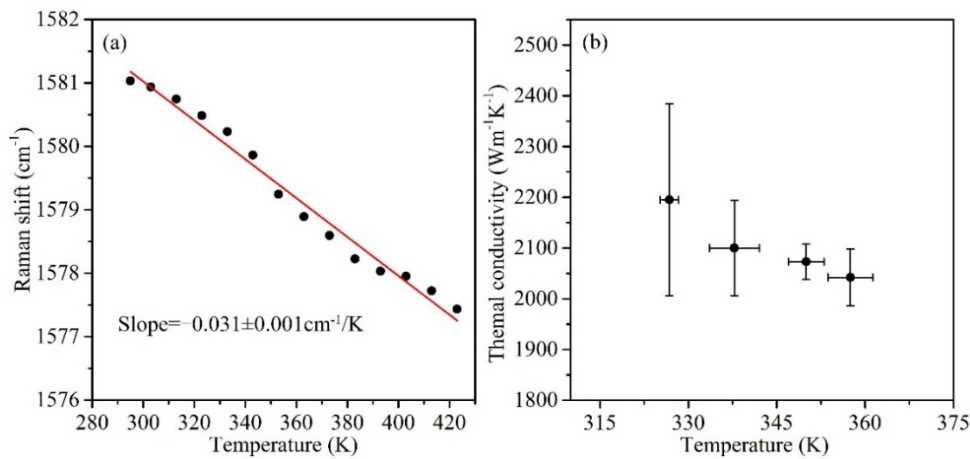

**Fig. S7.** (a) G band frequency of a monolayer graphene as a function of temperature, along with the corresponding linear fitting; (b) thermal conductivity of the graphene as a function of the Raman measured temperature.

## 8. Thermal equilibration on MD simulations using LAMMPS

An initial equilibration using NVT ensemble on the systems was followed by the calculation of the thermal conductivity at the NVE ensemble as discussed in the main text.

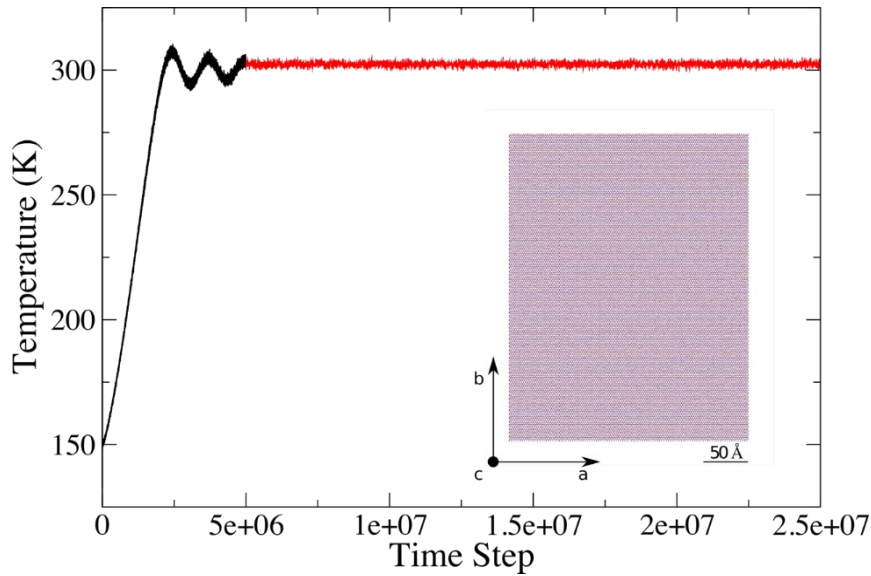

**Fig. S8.** Temperature versus time step for monolayer BN. The black curve shows the initial equilibration using NVT ensemble, and the red curve is the subsequent evolution in NVE. A time step of 0.5 fs was used. See Methods for details.

## 9. Thermal expansion coefficient (TEC) of SiO₂/Si substrate simulated by finite element method (FEM)

Fig. S9a shows the geometry model in Abaqus. The model was symmetric in both x-axis and y-axis. Only ¼ of the model was calculated for computational time efficiency. Fig. S9b shows the dimension of the model projected on the x-y plane. The 90 nm top sheet was $SiO_2$ while the 810 nm bottom represented Si. Different thickness ratios between $SiO_2$ and Si were tested and numerically proved that the ratio of 9 was both computational accuracy and efficiency. Additionally, 8 nodes linear brick was used in this study. Element size was set to 30 nm. Different element sizes were also tested to prove accuracy. The simulation was conducted with Abaqus Standard implicit solver. The strain distribution of $SiO_2$/Si (Fig. S9c) indicated the TEC of the top layer of 90 nm $SiO_2$/Si was ~60% of the TEC of Si, neither equal to the TEC of Si nor to that of $SiO_2$.

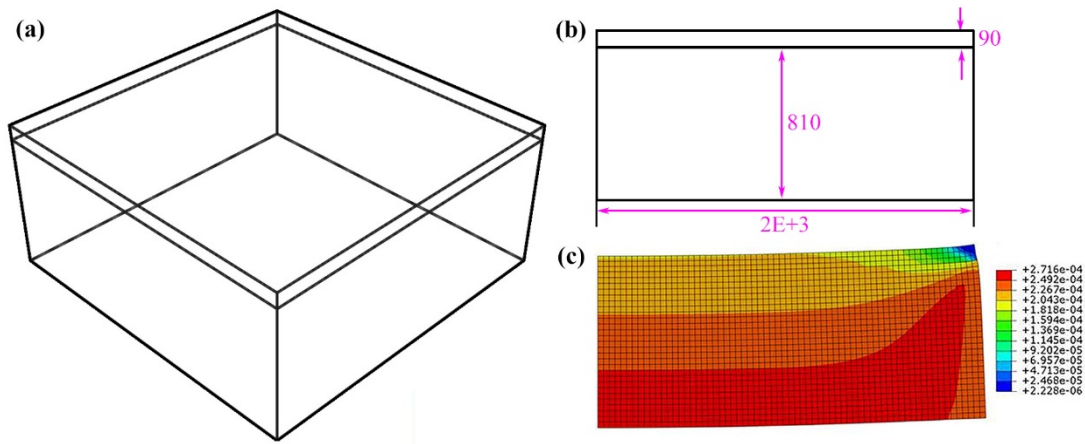

**Fig. S9.** (a) Geometry model in finite element software Abaqus; (b) Dimension of modeled sheets (unit: nm); (c) strain distribution of 90 nm $SiO_2$/Si.

## 9. Comparison of the thermal expansion coefficient of common 2D materials

The thermal expansion coefficients of some common 2D materials at 300 K are compared in Table S1, with the values of BN from this study.

**Table S1.** Thermal expansion coefficient of 2D materials ($10^{-6}$ $K^{-1}$)

| 2D materials | Monolayer | Bilayer | Trilayer |
|---|---|---|---|
| BN | -3.58±0.18 | -2.55±0.28 | -1.67±0.20 |
| Graphene | -21.4±3.7 | -10.9±2.5 | -8.7±1.7 |
| $MoS_2$ | 64.9±7.5 | 36.0±4.7 | 18.2±2.5 |
| $MoSe_2$ | 106.2±6.4 | 54.4±3.5 | 34.6±2.8 |
| $WS_2$ | 152.1±13.8 | 22.6±2.0 | 13.1±1.0 |
| $WSe_2$ | 154.2±6.9 | 41.8±2.5 | 27.4±2.9 |